\documentclass[12pt,preprint]{aastex}

\slugcomment{AJ, in press}

\shorttitle{NEOSurvey 1: Initial results}
\shortauthors{Trilling et al.}

\begin{document}


\title{NEOSurvey 1: Initial results from the Warm Spitzer
Exploration Science Survey of Near Earth Object Properties}


\author{David E. Trilling\altaffilmark{1,2,3}}

\author{Michael Mommert\altaffilmark{1}}

\author{Joseph Hora\altaffilmark{4}}

\author{Steve Chesley\altaffilmark{5}}

\author{Joshua Emery\altaffilmark{6}}

\author{Giovanni Fazio\altaffilmark{4}}

\author{Alan Harris\altaffilmark{7}}

\author{Michael Mueller\altaffilmark{8,9}}

\and

\author{Howard Smith\altaffilmark{4}}

\altaffiltext{1}{Northern Arizona University}
\altaffiltext{2}{Visiting Scientist, South African Astronomical Observatory}
\altaffiltext{3}{Visiting Professor, University of the Western Cape}
\altaffiltext{4}{Harvard-Smithsonian Center for Astrophysics}
\altaffiltext{5}{Jet Propulsion Laboratory, California Institute of Technology}
\altaffiltext{6}{University of Tennessee, Knoxville}
\altaffiltext{7}{DLR/German Aerospace Center}
\altaffiltext{8}{Netherlands Institute for Space Research}
\altaffiltext{9}{Kapteyn Astronomical Institute}



\begin{abstract}
Near Earth Objects (NEOs) are small Solar System bodies whose orbits
bring them close to the Earth's orbit.
We are carrying out
a Warm Spitzer Cycle~11 Exploration Science program entitled
NEOSurvey --- 
a fast and efficient flux-limited survey of 597~known
NEOs in which we derive diameter and albedo for each target.
The vast majority of our targets are too faint to be 
observed by NEOWISE, though a small sample has
been or will
be observed by both observatories, which allows
for a cross-check of our mutual results.
Our primary goal is to create a large
and uniform catalog of NEO properties.
We present here the first results from this new program:
fluxes and derived diameters and albedos for 80~NEOs,
together with a description of the overall program and approach,
including several updates to our thermal model.
The largest source of error in our diameter and albedo
solutions, which derive from our 
single band thermal emission measurements,
is uncertainty in $\eta$, the
beaming parameter used in our thermal modeling;
for albedos, improvements in Solar System absolute
magnitudes would also help significantly.
All data and derived diameters and albedos from
this entire program are being posted
on a publicly accessible webpage at {\tt nearearthobjects.nau.edu} .
\end{abstract}





\section{Introduction}

Near Earth Objects (NEOs) are small Solar System
bodies whose orbits bring them close to the Earth's
orbit. NEOs lie at the intersection of 
science, space exploration, and civil defense.
Because NEOs are constantly being replenished from
sources elsewhere in the Solar System,
they act as 
compositional and
dynamical tracers and allow
us to probe
environmental conditions throughout the Solar System
and the history of our planetary system.
NEOs are the parent bodies of
meteorites, one of our key sources of detailed knowledge about the Solar
System's development, and
NEO studies provide the needed context for this work.
The space exploration of NEOs is
carried out both through
robotic spacecraft 
such as NEAR
\citep{veverka};
Hayabusa \citep{fujiwara};
Chang'e~2 \citep{ji};
Hayabusa~2 \citep{abe}; and
OSIRIS-REx \citep{lauretta}; as well as
remote sensing observations
using IRAS, Spitzer, WISE/NEOWISE, Akari, and other 
assets (as reviewed in \citealt{asteroidsiv}).
Energetically,
some NEOs are easier to reach with spacecraft than the Earth's
moon \citep{barbee},
and NEOs offer a large number of targets with a range of physical
properties and histories. NEOs present advanced 
astronautical challenges, and President Obama announced
in 2010 that missions to asteroids
would form the cornerstone of NASA's exploration for the
next decade or more; the new
Asteroid Retrieval Mission
\citep{abell2016}
is
part of those activities.
Finally, NEOs are a civil defense matter:
the impact threat from NEOs is real,
as demonstrated again recently in Chelyabinsk, Russia, in February,
2013
\citep{popova,brown2013}.
Understanding the
number and properties of NEOs affects our planning
strategies, international cooperation, and
overall risk assessment.

The Spitzer Space Telescope is a powerful
NEO characterization system.
NEOs typically have daytime temperatures 
around 300~K. Hence, their thermal
emission at 4.5~microns is almost always significantly
larger than their reflected light at that wavelength. We can therefore
employ a thermal model to derive NEO properties, with the
primary result being derived diameters and albedos
\citep{paper1}.
Measuring the size distribution and albedos
and compositions for a large fraction of all known
NEOs allows
us to understand the scientific, exploration, and
civil defense-related properties of the NEO
population.
Spitzer's sensitivity at 4.5~microns
is unparalleled, allowing us to observe many
hundreds of NEOs over the course of the 20~month
Cycle~11 period.
%

We present here a new Warm Spitzer Exploration Science program
entitled NEOSurvey: An Exploration Science Survey of Near
Earth Object Properties (PID11002; PI: Trilling). In this paper
we present the design of our program and results
for the first 80~NEOs observed in this program; results
for the complete sample will appear in a forthcoming paper.
We also introduce {\tt nearearthobjects.nau.edu}, a publicly-accessible
web page where all data and results from this program are
posted, typically within a week of the data being released
by the Spitzer Science Center.

In the 710~hours assigned to this program we
use Spitzer to observe 597~NEOs; the vast majority 
of these are
not accessible with any other facility.
For each observed target, we use a thermal
model to derive the diameter and albedo.
This sample, when combined with 
existing data from our previous ExploreNEOs work
(Spitzer Cycle~6+7 Exploration Science program),
together with results from NEOWISE and Akari,
includes nearly 20\% of all NEOs known at the
inception of this project.
The majority of the infrared measurements of NEOs in this combined
catalog with 
$H>21$ --- bodies smaller than
around 175~meters diameter ---
will come from this new program.

\section{Project design}


Our technical approach has been honed through 
observations of more than 600~NEOs with Spitzer
in our Cycles~6+7 ExploreNEOs program
\citep{paper1,paper6,paper2,paper3,paper5,paper8}, as well
as through a number of Spitzer programs to study
individual NEOs of interest
\citep[e.g.,][]{mueller2007,mommert2014b,mommert2014c}.
The observation design has been proven to deliver high quality
photometry. The sensitivity estimates we use here
are empirically
derived from our many measurements of NEOs, and are
more appropriate for observations of moving targets
than the estimates provided by Spitzer's
sensitivity estimating tool
SENS-PET\footnote{{\tt http://ssc.spitzer.caltech.edu/warmmission/propkit/pet/senspet/}}.
All of the technical steps are described in detail
in our various papers,
including 
our seminal ExploreNEOs paper,
\citet{paper1}.

\subsection{Target selection and observation planning}

We selected our sample targets based on Spitzer Cycle~11 observability
from the 24-Oct-2014 version of the Minor Planet Center NEO
list\footnote{{\tt http://www.minorplanetcenter.net/iau/MPCORB/NEA.txt}}. Observability
was checked using the NASA Horizons
system\footnote{{\tt http://ssd.jpl.nasa.gov/horizons.cgi}} for each day in
Cycle~11 (February, 2015, through September, 2016);
Horizons was accessed with the 
{\tt Callhorizons}\footnote{{\tt https://pypi.python.org/pypi/CALLHORIZONS}} Python module, 
which retrieves ephemerides and orbital elements
from the Horizons system based on user queries.
An object was considered a potential target when 
its solar elongation with respect to Spitzer complied with the observatory constraints (82.5--120~degrees), its
positional uncertainty was less than $150$~arcsec at 3$\sigma$
(so that the target would fall within the detector
FOV), it had
a proper motion of less than $1$~arcsec/s (though 
in practice no objects are excluded by this
constraint),
and its distance from the
Galactic plane was larger than $15^{\circ}$. These requirements are
based on experience from our successful ExploreNEOs program
\citep{paper1}.
From this list of potential observations and targets,
we removed all targets that had already been observed by Spitzer
(ExploreNEOs: \citealt{paper6}), WISE
\citep{mainzer2011b,mainzer2012,mainzer2014a}, or
Akari
\citep{usui,hasegawa}.

\begin{figure}
\includegraphics{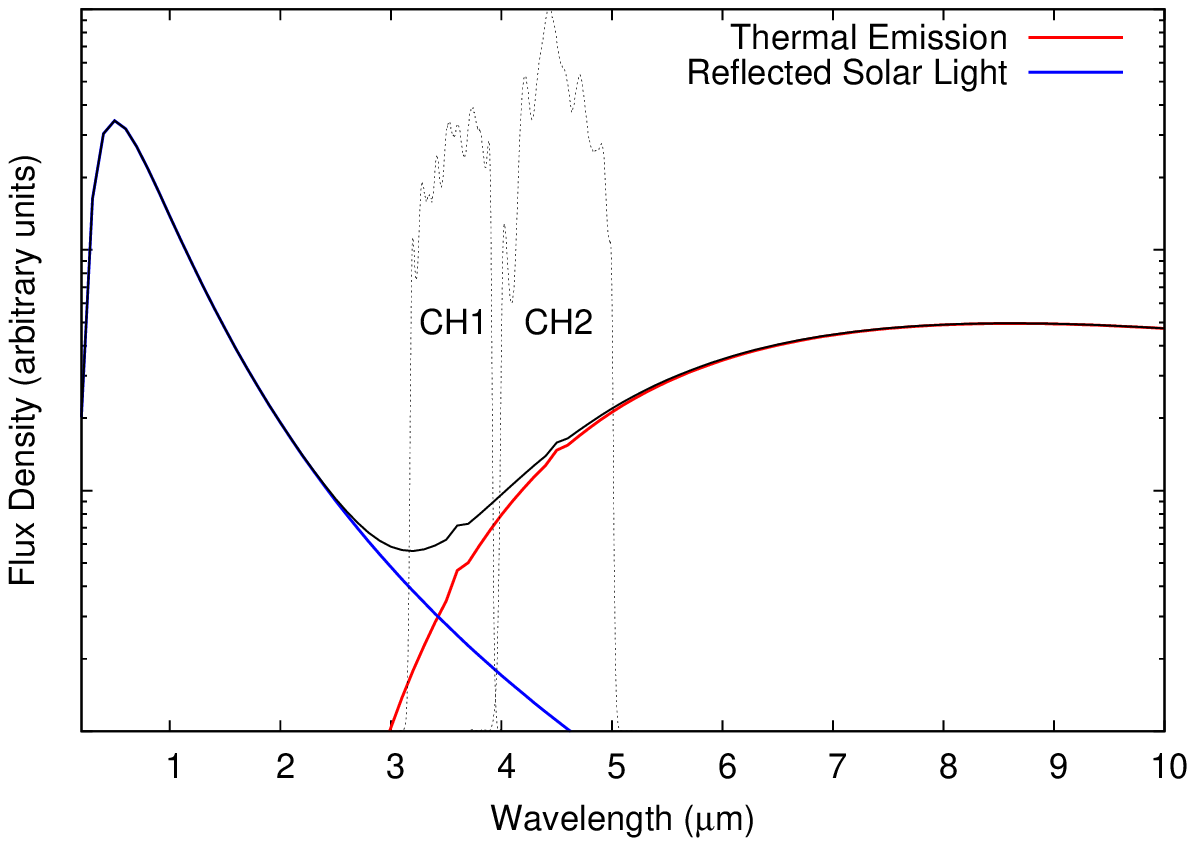}
\caption{Spectral energy distribution of a
typical arbitrary
NEO with
mean surface temperature of 300~K
and subsolar temperature of 360~K,
which are typical temperatures for objects
in this program.
The black line (total
observed radiation) is the sum of reflected
light (blue line) and thermal emission (red line). In IRAC CH1,
reflected light and thermal emission can be
comparable, depending on albedo
and heliocentric distance,
making interpretation difficult.
In IRAC CH2, thermal emission dominates, and
that measured flux can readily be used for thermal modeling.
The IRAC bandpasses are also shown for reference.}
\label{sed}
\end{figure}

Spitzer's Infrared Array Camera (IRAC; \citealt{fazio})
is the only instrument available in Spitzer's Warm Mission,
providing imaging
at 3.6~and 4.5~microns (CH1 and CH2,
respectively). The field of view is 5.2~arcmin.
Fluxes from NEOs at CH2 are always
dominated by thermal emission
(Figure~\ref{sed})
but the ratio of reflected light to thermal
emission in CH1 cannot be determined
if the albedo at 3.6~microns is not known
\citep{mainzer2011b}.
In our previous ExploreNEOs program
\citep{paper6}
we found that 
CH1 observations are generally not
useful for thermal modeling.
We therefore do not observe in CH1 in
this program, saving more than 50\% on
our total time request\footnote{
For (very) low albedo objects,
the thermal emission in CH1 
at 3.6~microns can be significant,
and in those cases both CH1 and CH2
can be used for thermal modeling
\citep{paper1,mainzer2012,nugent2015}, and
indeed thermal emission for some (very)
low albedo objects can be measured from
the ground in near-infrared 
spectra \citep{rivkin2005}. However, since
we do not have any {\em a priori} knowledge
of which objects in our sample might have
sufficiently low albedos for CH1
measurements to be useful for thermal
modeling, and since that number is 
likely to be small \citep{paper1}, we choose
to observe only in CH2 in this program.}.

For each potential target we predicted the thermal-infrared flux density
in IRAC CH2
as a function of time.
Our flux predictions were based on the absolute
optical magnitude $H$, a measure of $D^2 p_V$, as reported by
Horizons ($D$ is diameter and $p_V$ is albedo).
$H$ values for NEOs are of notoriously low quality and tend
to be skewed to brighter magnitudes
\citep{juric,rt2005}.
To account for this, we assumed an $H$ offset ($\Delta H$) of [0.6, 0.3,
0.0]~mag for [low, nominal, high] fluxes, respectively. Optical
fluxes were calculated from $H + \Delta H$ together with the observing
geometry. In our observation planning we assumed that
asteroids
are 
1.4~times more reflective at 4.5~microns than
in the $V$ band
\citep{trilling2008,paper2,mainzer2011b},
although in our analysis pipeline
we use an updated value that is derived
from NEOWISE data, as described below.
Thermal fluxes also depend on $p_V$ ($D$ is determined
from $H$ and $p_V$) and $\eta$ (see Section~\ref{modeling}). We assumed $p_V$ = [0.4, 0.2, 0.05]
for [low, nominal, high] thermal fluxes. The nominal $\eta$ value was
determined from the solar phase angle $\alpha$ using the linear
relation given by \citet{wolters}, which is generally in
agreement with the newer results of \citet{mainzer2011b}
and the approach used in this work 
(see \S\ref{beaming}, below);
0.3~was
[added, subtracted] for [low, high] fluxes to capture the scatter in
the empirical relationship derived in \citet{wolters}.
From these parameters, we calculated the predicted
thermal fluxes using the Near Earth Asteroid
Thermal Model \citep[discussed further in Section~\ref{modeling}]{harris1998}.
We calculated [low, nominal, high]
predicted fluxes in the 4.5~$\mu$m IRAC band for each day of
Cycle 11. The resulting NEATM fluxes were convolved with the CH2
passband to yield ``color-corrected'' in-band
fluxes. 

After removing all dates 
where an NEO's {\em high} predicted
flux could saturate the detector (using
saturation values from the IRAC
Warm Mission Handbook),
we identified a five day window centered
on the peak brightness during Cycle~11.
We used the minimum flux calculated
for the {\em low} predicted values
in this five day window to calculate our integration
times and build our AORs. 
This five day window dictates
the timing constraint that we place on the AOR.
A window of this size
allows good scheduling flexibility
while allowing us to use the shortest possible integration
times.
Using the {\em low} prediction
values here ensures that our SNR requirement
will be met or exceeded.
Our experience has shown that the ratio of
observed to {\em low} predicted flux is
in the range~1--10, so our conservative
prediction technique is appropriate.

Our photometric requirement is SNR$\geq$15,
at which point, our detailed thermal
modeling shows,
the flux density uncertainty becomes
small compared to overall model uncertainties.
We have empirically measured the Warm Spitzer
CH2 sensitivity (5$\sigma$)
for
faint moving objects to be
2.2~$\mu$Jy in 5~hours and
1.8~$\mu$Jy in 10~hours, scaling
roughly as the square root of integration
time for shorter integrations 
(Mommert et al.\ 2014b).
This is less sensitive
than predicted by SENS-PET due to 
contamination from background sources
that cannot be 
completely subtracted out
in the moving frame.

\begin{figure}
\includegraphics{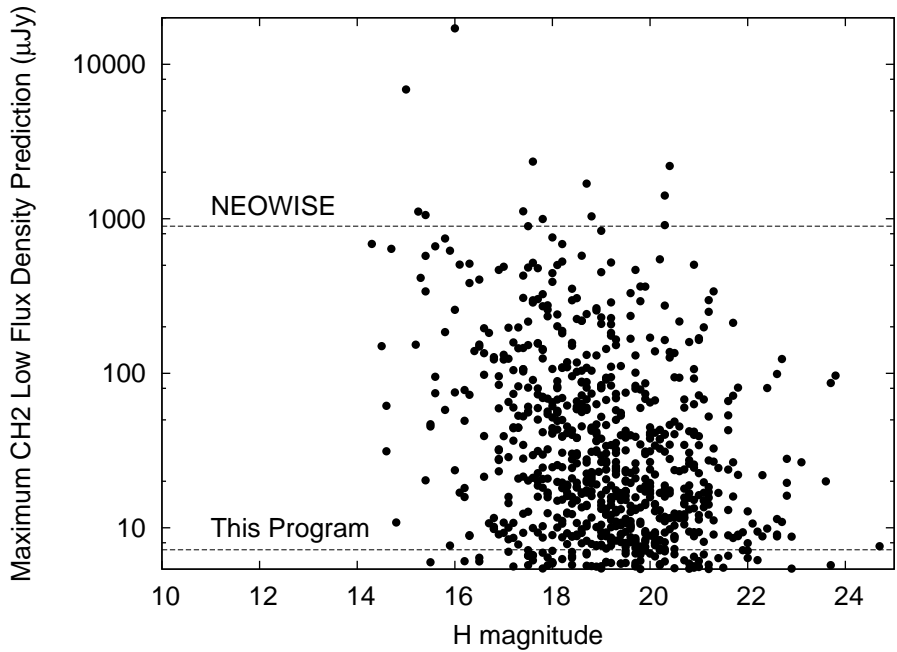}
\caption{
Maximum {\em low} CH2 flux density predictions as a function of $H$
magnitude.
The dots show all NEOs accessible with Spitzer in Cycle~11.
Dashed lines indicate the $15\sigma$
sensitivities
for the NEOWISE survey and this program.
The sensitivity limit for NEOWISE W2 (890~$\mu$Jy)
is derived from flux densities tabulated in Mainzer et al.\ (2014b) and the
flux calibration information in Wright et al.\ (2010).
We set our integration cutoff to be 10,000~seconds,
and 
observe all targets whose peak brightnesses
are greater than 7.2~$\mu$Jy, as indicated: all dots above
the dotted line marked ``this program.''
There are many small NEOs in this sample,
with some 200~targets smaller than 200~meters
($H>20$).
The vast majority of our targets are too faint to be 
observed by NEOWISE. 
Objects above the NEOWISE flux limit as 
seen by Spitzer may not be bright enough to 
be detected by NEOWISE when in the NEOWISE
field of view, and the reverse: some objects that
are below the NEOWISE faint limit as seen by Spitzer
may be brighter when they are observed by
NEOWISE. Therefore, 
a small number of NEOs will
be observed by both Spitzer and NEOWISE, which allows
for a cross-check of our mutual results.
\label{comparison}
}
\end{figure}

Our threshold is set to integration
times of $\leq$10,000~seconds (2.7~hours), which
produces an AOR clock time including overheads
of 3.2~hours.
All sample targets have a 4.5~$\mu$m thermal-infrared
flux density of 7.2~$\mu$Jy or greater.
In 710.1~hours of total observation time, including all
overheads, we
will observe 597~NEOs.
Our final sample is shown in Figures~\ref{comparison}
and~\ref{hfig}.

\subsection{Observing strategy}

We use a standard moving single target AOR,
in which the telescope tracks at the appropriate
moving object rates for the object,
and a medium cycling dither,
with only the 4.5~micron field integrating on the NEO
(as we did recently for 2011~MD \citep{mommert2014c}).
As described below, the 4.5~micron band provides the most information
on the thermal emission from the NEO, whereas the 3.6~micron band can
have a significant (and in general
unknown) contribution from reflected light and does not significantly improve the derivation of an albedo and size.  
Not observing in CH1 yields significant time savings
as well; since most NEOs are fainter in CH1 than
in CH2 (Figure~\ref{sed}), observing in CH1
would more than double the needed time.
Each target was assigned to one of six uniform observing
templates with total frame times [120, 480, 1200, 2400, 6000, 10000]~seconds.
The individual frame times are 12~sec and 30~sec for the first two
bins, respectively, and 100~sec for the remainder; this allows many
dithers for each observation.

The frame times used for each object were chosen based on our
estimate of the source flux in order to obtain the required sensitivity
using the minimum total observation time. However, we have seen that
occasionally the NEO is brighter than expected, possibly due to errors
in the previously determined $H$ magnitude, or as in the case of Don
Quixote, the result of cometary activity \citep{mommert2014a}.
We therefore use HDR mode for the 12~and 30~second
frame times, which minimizes 
the chance of the object saturating. However,
if saturation does occur,
we can use a PSF-fitting technique to recover the NEO flux, as 
we did 
for Don Quixote\footnote{For saturated
observations the Warm Mission PSF can be fit to the unsaturated
wings of the saturated NEO observation to determine the object flux
to better than 5\% accuracy \citep{marengo2009}.}.
(Using HDR mode 
for the brightest targets
increases the total overhead only a small amount.)
For the 
100~second frame times,
where the objects are very likely much fainter
than the saturation limit,
we do not use HDR mode.

We have also checked for objects that move only slightly
during the minimum observation time required to reach SNR$\geq$15. If
an object moves only a small distance during the AOR, it is difficult
to subtract the background field and perform accurate photometry,
especially if it falls in close proximity to a bright source.  We have
therefore found all objects that would move less than $\sim$3~FWHMs during
the observation, and added additional integrations for these sources
so that the object moves sufficiently to be able to
measure and subtract the background field. This has been done for
approximately 50~sources, which
are all
relatively bright targets and have short duration AORs.

The objects in our target pool are distributed all
over the sky and have an essentially 
uniform distribution of ecliptic longitudes
(except at the Galactic plane crossings).
No AOR is longer than 3.2~hours, including
all overheads,
and more than half the targets have AORs
that are 30~minutes or less.
As such, 
this program maximizes both total NEO yield
and efficiency.

\subsection{Data processing}

The data processing is performed in a manner similar to our
previous reductions for
ExploreNEOs \citep{paper1} and our other observations of
NEOs \citep{mommert2014b,mommert2014c}.  The BCDs are downloaded from
the Spitzer archive, and we use {\tt IRACproc/mopex} in the moving object mode to mosaic the frames relative to the NEO position.  For objects that are relatively bright compared to the background, and for low background fields, the outlier rejection process in the mosaicking software is sufficient to remove background objects.  For fainter objects or regions of relatively high background, we often must subtract the background field before constructing the NEO mosaic.  We first mosaic the data in the standard (non-moving object) mode to construct an image of the background field (the NEO is masked from the individual frames before making the background mosaic).  Then we subtract the background image from each BCD, and mask any remaining strong residuals from bright background objects.  We then construct the NEO mosaic from the background-subtracted BCDs.  Once the NEO mosaic has been constructed, we perform aperture photometry to measure the NEO flux.  We use an aperture correction derived from observations of standard stars during the Warm Mission to determine the flux.

Each mosaic is examined during the photometry process to ensure that there are no image artifacts present that will affect the photometry, and
that the NEO is found in the field and not confused with background objects. 
The SSC produces a data delivery
approximately every two weeks that will contain $\sim$14 new NEO
observations from the campaign that ended two weeks prior.
In most cases, we can extract fluxes and perform
thermal modeling within a few days
of data delivery.

\section{Thermal modeling \label{modeling}}

In order to interpret the thermal-infrared flux densities measured by
Spitzer, we utilize the Near-Earth Asteroid Thermal Model
\citep[NEATM,][]{harris1998}. The NEATM determines the
thermal-infrared flux density emitted from the surface of an asteroid
by integrating the Planck function over the body's surface temperature
distribution. By combining thermal-infrared observations and optical
brightness measurements ($H$), NEATM is used to fit the model flux
density to the measurement in order to independently derive $D$ and
$p_V$ of the target body. The NEATM incorporates a variable adjustment
to the model surface temperature through the beaming parameter $\eta$,
allowing for a correction for thermal effects of shape, spin state,
thermal inertia, and surface roughness, and enabling the model and
observed thermal continua to be accurately matched.

NEATM has been applied in the successful ExploreNEOs \citep{paper1}
and NEOWISE \citep{mainzer2011b} programs, as well as for other
asteroid observations, to measure the physical properties of a large
number of NEOs.  
In contrast,
more detailed
thermophysical modeling
\citep{delboasteroidsiv}
would require knowledge of shape and spin
state, which is generally unavailable for our poorly studied targets.
Even the hybrid NESTM
\citep{nestm}
requires assumptions about or default values 
of thermal inertia and rotation
period, neither of which is generally known for our targets.
Additionally,
\citet{nestm}
show that
NEATM appears to overestimate diameters
at phase angles larger than 45~degrees, but only by a significant
amount (more than 15\%) for phase angles larger than 
80~degrees. Only a few percent of our targets are observed
at phase angles this large.
Furthemore,
ultimately the \citet{nestm} approach is irrelevant
in our case because they
considered only the cases of floating $\eta$ or fixed $\eta$. 
Our use of $\eta$ as a linear function of 
$\alpha$ will result in different behavior, and
\citet{paper2} found
no evidence in our approach for a 
significant phase-angle-dependent bias 
as a result of neglecting night-side
thermal emission (their Figures~4 and~5).
We therefore use NEATM, as
it adequately explains our data and does not limit
our analysis or interpretation.

In our (previous) ExploreNEOs program we fit the NEATM to
$H$ and single-band infrared data \citep[CH2 only, since CH1 data are heavily
contaminated with reflected solar light;][]{paper3}, deriving
robust results: \citet{paper2} found that this technique leads to
diameters and albedos that are accurate to within 20\% and 50\%,
respectively,
compared to previously published spacecraft, radar, and radiometric-based 
solutions.
In this program, we apply the same
modeling technique, which uses a value of $\eta$ that 
is not fit independently to the data but rather
is based on a
linear relation between $\eta$ and the solar phase angle $\alpha$, but
with a more sophisticated statistical analysis of the
uncertainties. In the following subsections, we provide a brief introduction to
the thermal modeling pipeline and elaborate on concepts of the pipeline
that differ from the ExploreNEOs approach \citep{paper1, paper3}.



\subsection{Reflected Solar Light}

The measured flux density includes both thermal emission from the
asteroid's surface and reflected solar light.
The latter is generally small, but in some cases
may be significant.
We subtract
this contribution from the measured flux densites using the method
described by \citet{paper3}, but with an updated
reflectance ratio between the optical and IR wavelengths of
1.6$\pm$0.3, which is based on measurements by WISE
\citep{mainzer2011b}, who assumed, as do we, that NEO albedos 
are equal at 3.6~and 4.5~microns.
This albedo ratio can differ throughout the NEO population,
and our error bar on this ratio
captures this variance.
We subtract the (estimated) reflected 
light contribution before calculating the color 
correction because reflected sunlight has a 
different spectral shape than asteroid thermal
emission; the reflected light spectral shape is
very close to the one used in the IRAC
flux calibration, so the corresponding color corrections for
the reflected sunlight is negligible.

In our previous program,
ExploreNEOs, we corrected solar flux densities by converting
from IRAC in-band flux densities to monochromatic flux densities by
integrating over the model surface temperature distribution and the product
of the model spectral energy distribution and the IRAC CH2 bandpass
\citep{paper3}, requiring an iterative approach. By contrast, in the
current program, we
use a different approach in which we color-correct locally emitted
thermal flux density from monochromatic to IRAC CH2 in-band fluxes
using the exact model temperature of that surface element at the time the
Planck equation is integrated over the surface temperature
distribution. 
In other words, we derive separate color 
corrections for each spot on the surface, 
similar to the approach used in \citet{mainzer2011a}.
Using this approach, 
the color correction is properly iterated
as part of the fitting process.
The derived color corrections are in the range
5\% -- 20\%; the only uncertainty associated comes from
the ratio of reflectivity in the infrared compared to
the optical \citep{paper3}, and our error bars (described above) already
capture our lack of knowledge of this reflectance
ratio \citep{paper3}.
In the measured 4.5~micron flux, the reflected
light fraction is between
0.01\% and 29\%, with a mean
contribution of 1\% and a median
value of 0.7\%.
This new method produces color corrections
that agree at the 1--2\% level with our previous approach
\citep{paper3}.

\subsection{$H$ magnitudes \label{Hmag}}

We obtain our optical $H$ magnitudes from the ASTDyS\footnote{{\tt
    http://hamilton.dm.unipi.it/astdys/} --- see also \\
{\tt http://newton.dm.unipi.it/neodys/index.php?pc=7.2\#mag}} database and improve them
using statistical corrections provided by \citet{pravec2012} over the
given range of $H$ magnitudes; uncertainties are also adopted from
that work. For the range of $H > 20.0$, we apply a generic offset
of 0.3~mag (we reduce the brightness relative to the catalog) and
adopt an uncertainty of 0.3~mag, which is supported by
\citet{pravec2012}, \citet{juric}, and \citet{paper7}. In the
future we will replace ASTDyS optical data with more accurate
$H$ magnitudes from ground-based surveys, where available.


\subsection{Beaming Parameter Phase Relation \label{beaming}}

Because we model single-band IRAC data, we are unable to use $\eta$ as a
free-floating fitting parameter.
Instead, we have to make use of
existing NEO thermal-infrared data to derive a robust estimate for
$\eta$. 
As a result of asteroid surface roughness, thermal emission
from the surface can be preferentially
directed toward the Sun, which is referred to as
infrared beaming. 
Energy conservation therefore
requires that an observer at a high phase angle observes less thermal radiation than would be received from a smooth object, leading to a trend of increasing $\eta$ with solar phase
angle $\alpha$.
This trend has been quantified in the literature
\citep{delbo2003, wolters, mainzer2011b} using linear relations
between the parameters.
In general, all of ExploreNEOs, NEOSurvey, and Warm NEOWISE must
use such a derived (not fitted) $\eta$,
though NEOWISE was able
to fit $\eta$ directly for data obtained during their cryogenic
mission, where some NEOs were detected at multiple
thermal wavelengths, and both ExploreNEOs
and Warm NEOWISE can fit both CH1 and CH2 
thermal fluxes for very low albedo objects.

We also use a linear relation to 
derive the best-fit
$\eta$ based on its phase angle $\alpha$, but
here we take a more sophisticated approach to sample the
structure and full range in the $\eta$ distribution of NEOs. Based on
values of $\eta$ from NEOWISE \citep{mainzer2011b}, which provides the
largest published sample of fitted $\eta$ values, we derive a new relation.
Phase angles of the observed objects are derived by
querying the Minor Planet Center database for WISE observations and
represent average values. Figure \ref{fig:etarelation} (left) shows
the $\eta$ data, the literature $\eta$-$\alpha$ relations, and our
newly derived relation (see next paragraph).
The NEOWISE fractional uncertainties in $\eta$ are uniformly
distributed throughout this space.
The high degree of scatter in $\eta$ across
the whole range of phase angle is an indication of the heterogeneity
in the thermal properties of NEOs and is a reminder that a simple
linear relation is not sufficient to describe the variety in
$\eta$. Instead, the whole range of $\eta$ values must be taken into
account in a statistical approach to properly interpret
thermal-infrared data.  Here we derive a formalism
that provides the most likely $\eta$ as a function of the phase angle,
as well as an uncertainty distribution that samples the whole range of
possible $\eta$ values.

From Figure \ref{fig:etarelation} (left), we observe that there are no
NEOs with $\eta < 0.6$ or $\eta > 3.1$
(0.6~is the smallest value that $\eta$ can ever
take on \citep{mommert2012}).
Furthermore, there is a clear
grouping of low-$\eta$ NEOs but a lack of
low-$\eta$ NEOs at high phase angles. A comparison of the $\eta$
distributions derived by \citet{mainzer2011b} and \citet{wolters}
shows that the large population of NEOs with low $\eta$ is not a result of
the thermal-infrared nature of the NEOWISE survey, as it is also
present in the optically-selected sample by \citet{wolters}. For
the $\eta$ relation used in this work, we adopt a slope of 0.01~deg$^{-1}$, which
is in agreement with previous estimates and the slope of the lowest
$\eta$ values (we show below that a slightly different slope will
not affect the results). We choose the intercept of our linear
relation in such a way that it divides the sample of $\eta$ values in
Figure \ref{fig:etarelation} (left) in two equally-sized subsamples;
the probability to overestimate $\eta$ is the same as to underestimate
it.
Our new relation is 
$\eta(\alpha) =
  (0.01~{\rm deg}^{-1})\alpha + 0.87$.
We define the residual $\Delta \eta$ as the difference
between
the measured values of $\eta$ and a specific $\eta(\alpha)$. Figure
\ref{fig:etarelation} (right) shows that $\Delta \eta$ for the
individual $\eta(\alpha)$ relations agree well with one another; despite the
slightly different slopes, the shapes of the $\Delta \eta$ histograms
align.
We fit a log-normal distribution to the residuals using a
least-squares method. The fitted distribution has parameters
$\mu=0.655$ and $\sigma=0.628$, using the notation of
\citet{Limpert2001}.
The median of the $\Delta \eta$ distribution is 0.675.
The upper
and lower 1$\sigma$ confidence interval of $\eta$ uncertainties spans
the range from -0.3 to +0.55, relative to the nominal value of $\eta$
as derived by our $\eta(\alpha)$ relation (3$\sigma$ interval:~-0.55 to~+1.7).
Overall, the $\eta$-$\alpha$ parameter space shown in 
Figure~\ref{fig:etarelation} is poorly sampled,
and we assume that the averaged log-normal distribution shown
in the right panel of Figure~\ref{fig:etarelation} applies
across all phase angles. The data in hand do not warrant
or allow any more detailed treatment of these uncertainties.

\subsection{Uncertainties in derived diameter and albedo}

Uncertainties in diameter and albedo are derived in a Monte Carlo
approach very similar to the one described by \citet{paper3}. We
perform 10,000 simulations with randomized parameters, including the
measured flux densities that are varied based on the measured
uncertainties; the absolute magnitude $H$,
with Gaussian uncertainties from \citet{pravec2012}
or, for $H>20$, as a Gaussian with 1$\sigma$ uncertainties
of~0.3;
and the optical/IR
reflectance ratio, 
also varied as a Gaussian.
The beaming parameter $\eta$ is varied within the
log-normal distribution described above. 
Note that occasionally values $\eta > 3.1$ or $\eta < 0.6$
are drawn from the distribution; these extreme
values are rejected.
%
Clipping the $\eta$ distribution affects the final 
results in a negligible way and is in agreement with the
observations obtained by \citet{mainzer2011b}.

We
adopt the median of the diameter and albedo distributions
as our nominal solutions.
We derive
1$\sigma$ and 3$\sigma$ confidence intervals as those ranges of values
that bracket 68.3\% and 99.7\% of the values around the median,
respectively.
In Figure~\ref{uncertainties}
we show
representative cases of
the 
diameter and albedo probability distributions
that result from our thermal modeling.

The beaming parameter $\eta$ is 
not well constrained (Fig.~\ref{fig:etarelation}),
and drives the overall uncertainties in our model results.
For the results presented here,
more than 90\% of the total uncertainty
in diameter results from uncertainty in $\eta$. 
For albedo, 80\%--90\% of the total uncertainty
is due to $\eta$ uncertainties, with uncertainties
in $H$ responsible for most of the remaining uncertainty.
Uncertainties in
other aspects (such as emissivity) are about ten times smaller and are
therefore negligible.

Finally, we make no attempt to correct for
intrinsic NEO lightcurves for the targets in this program.
The elapsed time of our observations spans the 
range 10~minutes to 3.2~hours. For the longest exposures,
we might integrate over an entire asteroid rotation period,
as periods of seconds to a few hours hours are common for small
asteroids, in which case we derive the mean diameter
from our observations. For short exposures, we could easily
sample just one part of the lightcurve, in which case our
solution may represent the effective spherical diameter
that corresponds to a lightcurve minimum or maximum,
thereby under- or overestimating the true mean diameter.

We estimate how many of our observations are affected by lightcurve
effects with a Monte Carlo approach. Taking into account the measured lightcurve
amplitude distribution of
known NEOs \citep{warner},
we randomly sample sine curves with different amplitudes to simulate
the offset from the lightcurve-averaged thermal flux of the target in
our observations and produce diameter and albedo solutions
that correspond to this instantaneous sampling.
We find that
3--6\% of all trials have diameter discrepancies greater than our
quoted diameter uncertainty of 40\%. Hence, we expect the same
fraction of our targets to have diameters affected by
lightcurve-induced discrepancies greater than 1$\sigma$. 
For our albedo solutions,
5--9\%~of trials have solutions greater than
our quoted albedo uncertainty of 70\%.
If we smooth the lightcurve over 20\% of its period
(which is a more realistic approximation of
the situation),
we find that
fewer than 5\% of our diameters
and fewer than 9\% of our albedos
are affected by this lightcurve sampling.

Derived albedos that are (much) greater than
around~0.5 are suspicious, as such values are
rarely seen for asteroids for which albedos
have been reliably measured (for radar or spacecraft
measurements, for example, as opposed to radiometric
approaches), and are at odds with
expected surface compositions.
The most likely explanations for such
high albedos are $H$ magnitudes that are significantly
incorrect and/or (optical or thermal) lightcurve amplitudes and periods that
produce a high albedo solution. In both cases, additional
ground-based photometry can 
allow us to correct our albedo solution.
We have initiated such a program using a range
of ground-based telescopes, and in a future paper we
will report revised optical magnitudes and/or lightcurves
for these high albedo objects, along with re-calculated
albedos.
(Note that poor $H$~values and/or lightcurves
could also produce anomalously low
albedos, which may be harder to identify.
We will carry out ground-based
observations of the objects for which we 
find the lowest albedos to determine if
a correction is also needed for those
objects.)


\subsection{Validation of our approach}

A small number of our targets
will be
observed by NEOWISE --- it is impossible
to calculate which targets will overlap in the future,
since the NEOWISE
SPICE kernel (orbit) is only published a few months
ahead --- and we will use those targets
as cross-checks.
Furthermore, there are more than 100~NEOs
that have been observed by NEOWISE that
were also observed in our ExploreNEOs
program \citep{paper6}.
For these common objects,
we use our updated thermal model pipeline
and used only CH2 data (as if the ExploreNEOs
data were obtained in this NEOSurvey program)
to rederive the diameter and albedo from
our ExploreNEOs measurements, and compare
these to the published NEOWISE diameters and albedos
\citep{mainzer2011b}.
The mean ratio
of ExploreNEOs values to NEOWISE values
for [diameter, albedo]
is [0.9, 1.28].
The median value in this 
ratio of ExploreNEOs values to NEOWISE
values for [diameter, albedo]
is 
[0.9, 1.02].
This overall agreement, based on independent data measurements
and modeling, is quite good,
although we caution that for any individual object
significantly different solutions may exist.
For diameter,
70\% of 
the newly modeled ExploreNEOs
targets 
have 1$\sigma$~error bars that
overlap with NEOWISE values;
98\% of the targets agree at 3$\sigma$.
For albedo,
66\% of the newly modeled
ExploreNEOs targets 
have 1$\sigma$~error bars that
overlap with NEOWISE values;
98\% of the targets agree at 3$\sigma$.
This suggests that both the errors are largely
random, and that our error bar estimations
are appropriate.
The largest deviation is 
in albedo, where uncertainties in $H$~magnitude typically
introduce large errors, and where the two approaches
frequently use somewhat different $H$~magnitudes for their modeling.
In the future we will carry out a more comprehensive
comparison, considering targets that overlap with
both ExploreNEOs and NEOSurvey, and using common
$H$~values for the thermal modeling of the different thermal
measurements.

\section{First results from NEOSurvey}


We report here results for the first 80~NEOs
observed in this program. These NEOs were observed in February
and March,
2015. Table~\ref{datatable}
gives
the geometries,
measured fluxes, and 
derived
diameter and albedo results for these targets.
As described above, the
probability distributions for the results
are asymmetric and non-Gaussian. We therefore
provide 1$\sigma$ and 3$\sigma$ 
upper and lower diameter and albedo values for each target.


We introduce here {\tt nearearthobjects.nau.edu}, a publicly
accessible webpage where all information 
from this program is posted.
Table~\ref{webpage} lists the properties
for each observed object
that are available
in our online database,
where we list
complete observing
circumstances and modeling results.
The observing cadence for this program
is roughly one target per day. Data is made available
to us at a biweekly cadence. It takes us typically 1--2~days
from receiving the data to producing model results.
Therefore, our online database grows continuously,
in roughly two week intervals, over the duration of
this program. The latest version of our results
will always be available on that webpage.
Because of the Monte Carlo nature of our modeling
solutions, the values in this online table can
very slightly with time, but all such changes
are insignificant compared to the size of the
error bars.
The database ultimately will also host all ExploreNEOs data,
with results derived from the latest version of 
our pipeline.
Relevant publications will
also be posted at that webpage.

\section{Future work}

\begin{figure}
 \includegraphics{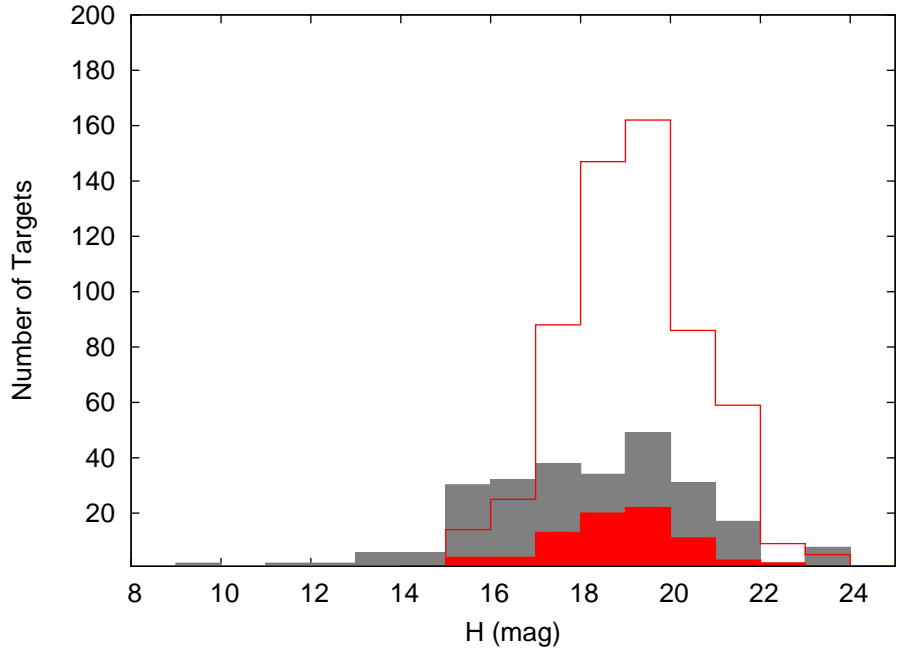}
\caption{Distribution of $H$ magnitudes 
for targets to be observed
in this sample (thin solid
red line and unshaded histogram) and by
NEOWISE
during Spitzer's Cycle~11 (shaded gray
region).
The 80~targets observed to date in this program
are shown as the shaded red histogram.
Because the NEOWISE SPICE kernel (orbit)
is only published a few months ahead,
it is impossible to calculate exactly which
NEOs will be observed by NEOWISE during
Spitzer's Cycle~11, but we can
closely estimate the $H$ distribution by
scaling the NEOWISE yield since reactivation
to the duration of Spitzer's Cycle~11.
A diameter of 100~meters corresponds
roughly to $H$~of~22.5.
At $H>20$, this project
contains more than three times as many
targets as NEOWISE and ExploreNEOs combined.
\label{hfig}
}
\end{figure}

\begin{figure}
\plotone{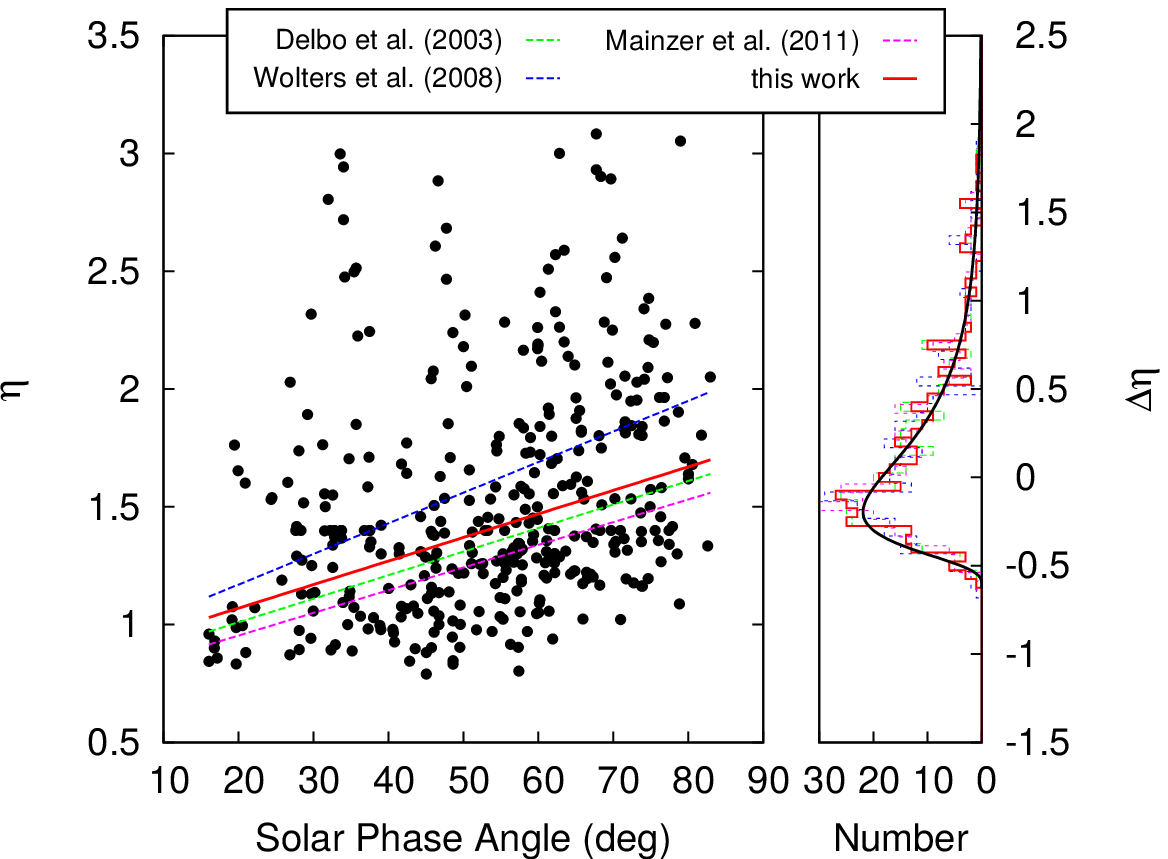}
\caption{{\bf Left}: Fitted $\eta$ values from \citet{mainzer2011b} as
  a function of the average solar phase angle $\alpha$. Dashed lines
  represent $\eta$-$\alpha$ relations from
the literature 
\citep{delbo2003,wolters,mainzer2011b},
and the red
  line indicates the relation used in this work, $\eta(\alpha) =
  (0.01~{\rm deg}^{-1})\alpha + 0.87$. Despite the fact that $\eta$ values
  scatter over the whole range from 0.6 to 3.1, a lack of
  high-$\alpha$ NEOs with low $\eta$, as well a general grouping of
  low-$\eta$ NEOs, suggest a trend between the two measures. {\bf
    Right}: Histograms of $\eta$ residuals; $\Delta \eta$ is defined
  as the difference between the value of each object in the left plot
  and the value of the respective $\eta$ relation at the object's
  $\alpha$. The histograms have been shifted in such a way that their
  median values coincide to improve readability. The black line is a
  log-normal distribution that has been fitted to
  the $\eta$-relation used in this work (see text).
}
\label{fig:etarelation}
\end{figure}

\begin{figure}
\plottwo{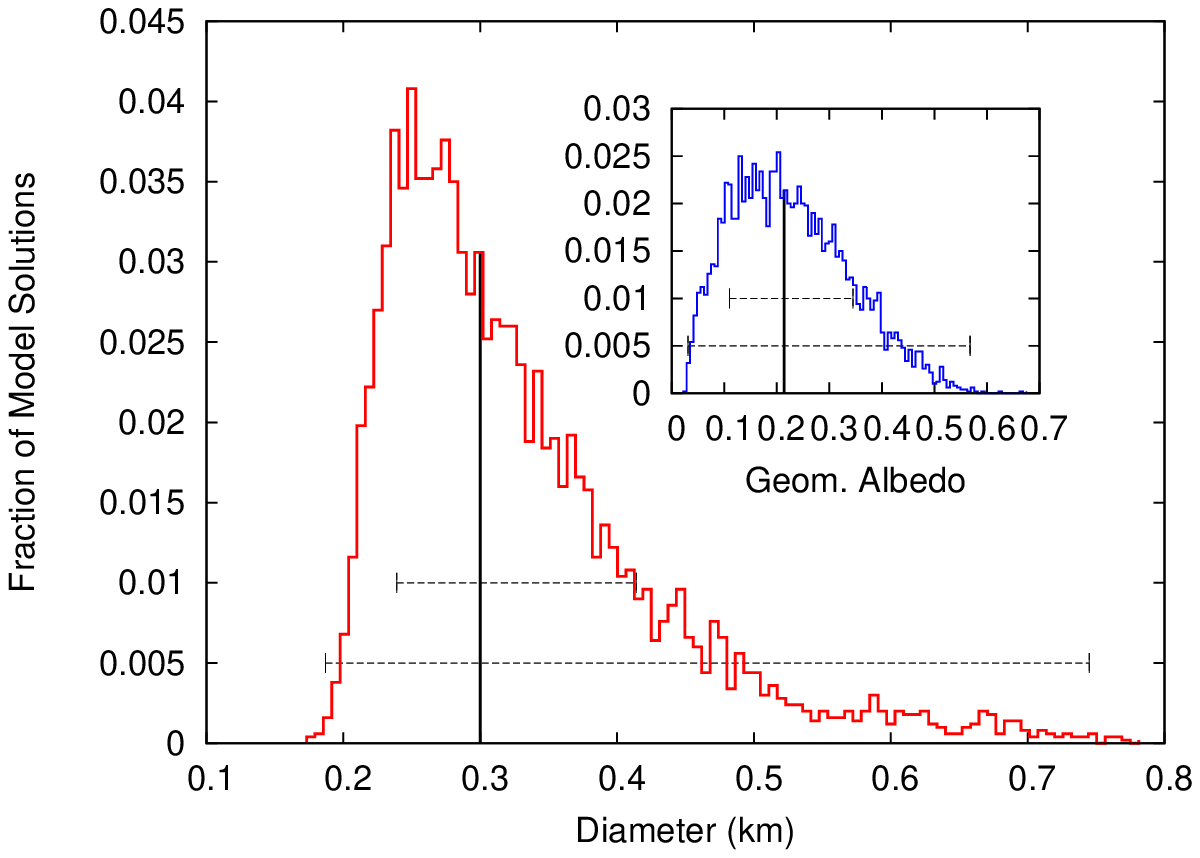}{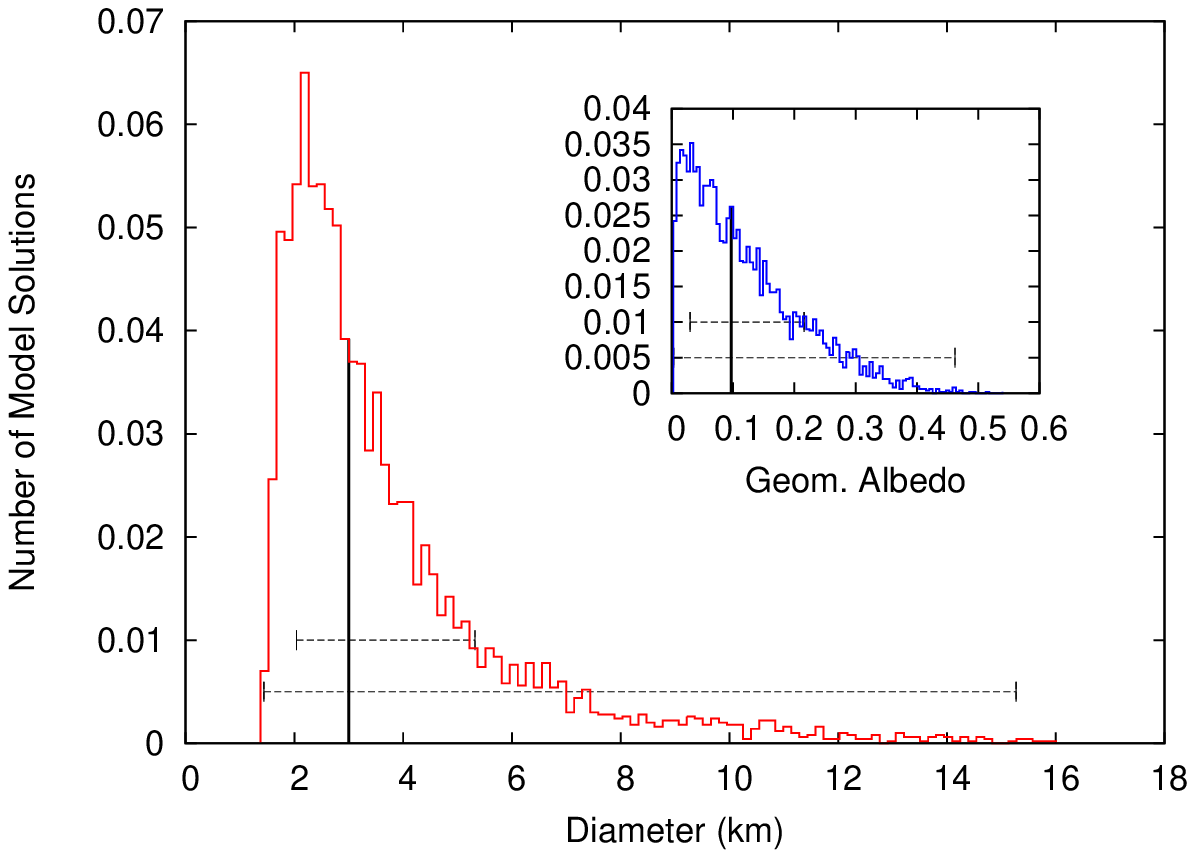}
\caption{Probability distributions for diameter and
albedo for the illustrative cases of 
2012~AD3
(left)
and
(8567) 1996 HW1
(right).
The vertical lines indicate our nominal
solutions (median of the probability distributions),
and 1$\sigma$ (encompassing
68.3\% of the trials surrounding
the median) and 3$\sigma$ 
(99.7\% of the trials) ranges are shown
with the horizontal error bars.
Both diameter and albedo distributions
are roughly log-normal.
In
this project we report both the 
1$\sigma$ and 3$\sigma$ ranges
for each parameter in order to capture the 
uncertainty distribution.
\label{uncertainties}}
\end{figure}

As our survey proceeds and our
catalog grows, a number of science investigations
will be enabled, including the following.
(1) Our sample is optically selected, so
any study of the properties 
of the entire NEO population requires 
rigorous debiasing, which is a signficant
challenge. Using debiasing approaches that we have 
developed \citep{paper6,paper8} and
continue to enhance, we
will 
measure the
size distribution of NEOs down to 100~meters.
We will
observe nearly 200~NEOs smaller than
200~meters ($H>20$; Figure~\ref{hfig}).
This will allow us to calculate an independent
measurement of the NEO size distribution, one
that is derived from a sample size at 100~meters
that is more than three times as large as
all the previous thermal infrared surveys combined.
(2) By applying
those same debiasing approaches
to the combination of
NEOSurvey, ExploreNEOs, and NEOWISE data
we will derive the distribution of albedos
and compositions
as a function of size. 
(3) 
We will search for ``dead comets'' --- bodies that
dynamically, and presumably compositionally, are of cometary origin,
but typically exhibit no cometary activity.
Our target sample includes 32~potential dead comets,
based on their low, comet-like Tisserand parameters, $T_J \leq 3.0$.
Our albedo measurements for all these objects may
provide further evidence for their cometary nature (cometary albedos
are extraordinarily low), as it did for
the NEO Don Quixote \citep{mommert2014a}.
(4) 
There are 109~targets in our sample
that obey the usual ``mission accessible'' definition
of $\Delta v<7$~km/s (and six targets
that have $\Delta v<5$~km/s).
We will measure diameters and
albedos for these
NEOs and,
together with historical temperature modeling as
in
\citet{paper3},
identify the most primitive 
objects --- and therefore most interesting for
mission planning --- in our sample.
(5) 
By determining the albedos of NEOs,
we can measure the compositions of small
bodies in the various source regions.
These results can be compared to models from
\citet{bottke2002}
and recent
work by \citet{granvik} to make connections among meteorite,
NEOs, and parent asteroids or asteroid families in the main belt,
thereby providing geologic, mineralogic, and Solar System context to
the objects studied in our laboratories and by our space missions.

The preceding list is not meant to be exhaustive, as many
other investigations will be possible, both by members
of the survey team and other researchers. The ultimate result
will be a catalog of some 2000~NEO properties, derived from
the union of NEOWISE, ExploreNEOs, Akari, and NEOSurvey results.
This will represent nearly 20\% of all known
NEOs in late~2016 when our NEOSurvey concludes.



\acknowledgments

We acknowledge the thorough and prompt hard work of the
staff at the Spitzer Science Center, without whom
the execution of this program would not be possible.
We also appreciate the support that
NASA's Planetary Science Division and 
Solar System Observations program provide
for the Spitzer mission and its NEO observations.
We thank two anonymous referees for their detailed
comments that improved this paper.
We also thank Scott Gaudi for useful advice.
Some of the computational analyses were run on Northern Arizona University's {\tt monsoon}
computing cluster, funded by Arizona's Technology and Research Initiative Fund. 
This work is based in part on observations made with the Spitzer Space
Telescope, which is operated by JPL/Caltech under a contract with
NASA. Support for this work was provided by NASA through an award
issued by JPL/Caltech. 
We make use of data from the AstDys database 
as well as extensive use of the JPL/Horizons system.



Facilities: \facility{Spitzer(IRAC)}.



\clearpage



\clearpage

\begin{deluxetable}{rlllllcccccc}
\tabletypesize{\scriptsize}
\rotate
\tablecaption{Target information and results \label{datatable}}
\tablewidth{0pt}
\tablehead{
\colhead{Target} & \colhead{r} & \colhead{$\Delta$} &
\colhead{phase} & \colhead{$H$} & \colhead{$\Delta H$} &
\colhead{F$_{4.5}$} & \colhead{$\eta$} & \colhead{$D$} & 
\colhead{$D~(3\sigma )$} & 
\colhead{$p_V$} & 
\colhead{$p_V~(3\sigma )$} \\
\colhead{} & \colhead{(AU)} & \colhead{(AU)} &
\colhead{(deg)} & \colhead{(mag)} & \colhead{(mag)} &
\colhead{($\mu$Jy)} & \colhead{} & \colhead{(m)} & 
\colhead{(m)} & 
\colhead{} & \colhead{} 
}
\startdata
8567 (1996 HW1) & 2.07 & 1.90 & 29.3 & 15.76 & 0.22 & 91$\pm$9 & 1.14 & 2900 (-900/+2400) & 1500---12100 & 0.10 (-0.07/+0.13) & 0.00---0.42 \\
35396 (1997 XF11) & 1.36 & 0.59 & 43.7 & 17.11 & 0.22 & 519$\pm$22 & 1.29 & 940 (-220/+480) & 380---1930 & 0.29 (-0.16/+0.21) & 0.03---0.60 \\
162117 (1998 SD15) & 1.25 & 0.46 & 50.2 & 19.41 & 0.22 & 147$\pm$11 & 1.34 & 350 (-80/+160) & 140---660 & 0.25 (-0.14/+0.18) & 0.02---0.49 \\
208023 (1999 AQ10) & 1.08 & 0.14 & 59.3 & 20.60 & 0.30 & 365$\pm$18 & 1.44 & 148 (-26/+52) & 52---201 & 0.46 (-0.22/+0.25) & 0.06---0.69 \\
(2004 TP1)  & 1.03 & 0.17 & 81.8 & 20.90 & 0.30 & 315$\pm$16 & 1.65 & 207 (-38/+74) & 70---253 & 0.18 (-0.09/+0.11) & 0.03---0.36 \\
390929 (2005 GP21) & 1.04 & 0.17 & 76.5 & 20.80 & 0.30 & 238$\pm$14 & 1.60 & 179 (-32/+62) & 62---222 & 0.26 (-0.12/+0.15) & 0.04---0.47 \\
434734 (2006 FX) & 1.05 & 0.36 & 75.4 & 20.30 & 0.30 & 66$\pm$8 & 1.59 & 198 (-35/+69) & 72---242 & 0.34 (-0.16/+0.19) & 0.06---0.55 \\
(2008 SU1)  & 1.19 & 0.63 & 58.6 & 19.51 & 0.22 & 113$\pm$10 & 1.42 & 410 (-90/+190) & 170---720 & 0.16 (-0.09/+0.11) & 0.02---0.36 \\
(2008 UF7)  & 1.41 & 0.62 & 40.7 & 19.41 & 0.22 & 54$\pm$7 & 1.25 & 330 (-80/+170) & 140---740 & 0.28 (-0.17/+0.22) & 0.02---0.65 \\
(2010 XP51)  & 1.24 & 0.39 & 48.1 & 18.93 & 0.22 & 2119$\pm$42 & 1.32 & 970 (-240/+510) & 430---2070 & 0.05 (-0.03/+0.04) & 0.00---0.14 \\
(2011 SM68)  & 1.09 & 0.50 & 68.3 & 19.89 & 0.22 & 57$\pm$7 & 1.51 & 241 (-44/+90) & 86---338 & 0.33 (-0.16/+0.18) & 0.05---0.51 \\
(2011 WS2)  & 1.52 & 1.22 & 41.8 & 17.59 & 0.22 & 149$\pm$11 & 1.26 & 1240 (-340/+750) & 590---3360 & 0.10 (-0.06/+0.10) & 0.01---0.33 \\
(2012 AD3)  & 1.12 & 0.47 & 64.7 & 19.89 & 0.22 & 96$\pm$9 & 1.48 & 290 (-60/+120) & 110---450 & 0.24 (-0.12/+0.14) & 0.03---0.43 \\
14827 Hypnos (1986 JK) & 1.32 & 0.93 & 50.2 & 18.65 & 0.22 & 65$\pm$7 & 1.35 & 520 (-120/+260) & 220---1030 & 0.22 (-0.12/+0.17) & 0.02---0.50 \\
4487 Pocahontas (1987 UA) & 1.22 & 0.66 & 56.4 & 17.68 & 0.22 & 789$\pm$26 & 1.41 & 1150 (-270/+570) & 480---2110 & 0.11 (-0.06/+0.09) & 0.01---0.29 \\
5332 Davidaguilar (1990 DA) & 1.64 & 1.32 & 38.2 & 15.09 & 0.22 & 634$\pm$24 & 1.22 & 3300 (-900/+2000) & 1500---9200 & 0.15 (-0.10/+0.14) & 0.01---0.44 \\
8034 Akka (1992 LR) & 1.41 & 0.78 & 44.8 & 18.16 & 0.22 & 78$\pm$8 & 1.29 & 540 (-120/+260) & 220---1170 & 0.33 (-0.18/+0.23) & 0.03---0.63 \\
32906 (1994 RH) & 1.33 & 0.66 & 48.2 & 16.15 & 0.22 & 2266$\pm$43 & 1.32 & 2080 (-510/+1100) & 880---4510 & 0.14 (-0.08/+0.11) & 0.01---0.36 \\
(1998 WP7)  & 1.07 & 0.35 & 71.6 & 20.18 & 0.22 & 133$\pm$11 & 1.55 & 256 (-49/+93) & 93---346 & 0.23 (-0.11/+0.13) & 0.04---0.38 \\
137078 (1998 XZ4) & 1.40 & 0.82 & 46.1 & 16.63 & 0.22 & 274$\pm$16 & 1.30 & 1070 (-230/+510) & 430---2140 & 0.34 (-0.19/+0.23) & 0.03---0.61 \\
53409 (1999 LU7) & 1.19 & 0.54 & 58.5 & 19.03 & 0.22 & 232$\pm$14 & 1.42 & 510 (-110/+240) & 200---870 & 0.17 (-0.09/+0.12) & 0.02---0.35 \\
102873 (1999 WK11) & 1.58 & 1.00 & 38.6 & 17.78 & 0.22 & 32$\pm$5 & 1.23 & 570 (-130/+290) & 240---1430 & 0.43 (-0.25/+0.31) & 0.03---0.85 \\
178601 (2000 CG59) & 1.48 & 1.21 & 42.8 & 17.88 & 0.22 & 90$\pm$9 & 1.26 & 920 (-240/+530) & 420---2340 & 0.15 (-0.09/+0.13) & 0.01---0.41 \\
159495 (2000 UV16) & 1.37 & 0.67 & 44.9 & 17.40 & 0.22 & 329$\pm$17 & 1.29 & 870 (-200/+440) & 350---1920 & 0.26 (-0.15/+0.19) & 0.02---0.53 \\
162723 (2000 VM2) & 1.21 & 0.35 & 48.2 & 17.68 & 0.22 & 812$\pm$27 & 1.32 & 600 (-120/+250) & 210---1040 & 0.42 (-0.21/+0.25) & 0.04---0.65 \\
162741 (2000 WG6) & 1.27 & 0.44 & 46.1 & 17.78 & 0.22 & 17336$\pm$191 & 1.30 & 3100 (-800/+1700) & 1500---7100 & 0.01 (-0.01/+0.01) & 0.00---0.04 \\
(2000 WP19)  & 1.04 & 0.06 & 68.1 & 22.70 & 0.30 & 2827$\pm$48 & 1.52 & 176 (-37/+73) & 68---261 & 0.05 (-0.02/+0.04) & 0.01---0.13 \\
68346 (2001 KZ66) & 1.35 & 0.95 & 48.9 & 17.11 & 0.22 & 219$\pm$14 & 1.32 & 1010 (-230/+520) & 400---2120 & 0.25 (-0.14/+0.17) & 0.03---0.51 \\
283460 (2001 PD1) & 1.21 & 0.75 & 56.5 & 18.55 & 0.22 & 230$\pm$14 & 1.40 & 700 (-160/+320) & 290---1300 & 0.14 (-0.07/+0.10) & 0.02---0.30 \\
317685 (2003 NO4) & 1.53 & 0.81 & 37.5 & 18.45 & 0.22 & 74$\pm$8 & 1.21 & 570 (-150/+320) & 260---1550 & 0.22 (-0.13/+0.19) & 0.01---0.57 \\
143992 (2004 AF) & 1.33 & 0.94 & 49.6 & 16.43 & 0.22 & 1064$\pm$30 & 1.33 & 2050 (-510/+1100) & 880---4440 & 0.11 (-0.07/+0.09) & 0.01---0.28 \\
267940 (2004 EM20) & 1.15 & 0.33 & 57.4 & 20.50 & 0.30 & 61$\pm$7 & 1.42 & 160 (-30/+62) & 58---245 & 0.44 (-0.21/+0.25) & 0.06---0.72 \\
154715 (2004 LB6) & 1.14 & 0.60 & 62.7 & 18.74 & 0.22 & 129$\pm$10 & 1.47 & 430 (-80/+170) & 150---650 & 0.30 (-0.15/+0.18) & 0.04---0.50 \\
(2004 QD14)  & 1.04 & 0.12 & 73.9 & 20.90 & 0.30 & 352$\pm$17 & 1.57 & 143 (-24/+49) & 47---174 & 0.37 (-0.18/+0.20) & 0.05---0.58 \\
318160 (2004 QZ2) & 1.14 & 0.47 & 62.4 & 18.36 & 0.22 & 256$\pm$15 & 1.46 & 490 (-90/+190) & 170---730 & 0.33 (-0.16/+0.19) & 0.04---0.53 \\
(2005 XY)  & 1.35 & 0.76 & 48.5 & 19.13 & 0.22 & 97$\pm$9 & 1.33 & 520 (-130/+270) & 230---1140 & 0.15 (-0.08/+0.12) & 0.02---0.38 \\
(2006 VD13)  & 1.09 & 0.31 & 68.0 & 19.32 & 0.22 & 360$\pm$18 & 1.53 & 370 (-70/+140) & 130---500 & 0.24 (-0.12/+0.14) & 0.03---0.41 \\
(2006 VD2)  & 1.10 & 0.41 & 67.6 & 19.51 & 0.22 & 314$\pm$16 & 1.51 & 430 (-90/+180) & 170---640 & 0.15 (-0.08/+0.09) & 0.02---0.28 \\
375103 (2007 TD71) & 1.30 & 0.70 & 50.7 & 18.74 & 0.22 & 91$\pm$9 & 1.34 & 460 (-100/+220) & 190---910 & 0.26 (-0.15/+0.18) & 0.02---0.53 \\
(2008 SQ)  & 1.25 & 0.51 & 52.1 & 19.80 & 0.22 & 86$\pm$9 & 1.36 & 300 (-70/+140) & 120---580 & 0.24 (-0.13/+0.17) & 0.02---0.49 \\
(2010 KU10)  & 1.15 & 0.40 & 60.5 & 19.80 & 0.22 & 236$\pm$14 & 1.44 & 360 (-80/+160) & 140---620 & 0.16 (-0.08/+0.11) & 0.02---0.34 \\
(2010 TJ7)  & 1.17 & 0.41 & 58.4 & 19.99 & 0.22 & 185$\pm$13 & 1.42 & 330 (-70/+150) & 130---570 & 0.17 (-0.09/+0.11) & 0.02---0.34 \\
(2010 VN65)  & 1.11 & 0.40 & 66.1 & 20.40 & 0.30 & 57$\pm$7 & 1.50 & 193 (-37/+72) & 70---281 & 0.32 (-0.16/+0.20) & 0.04---0.65 \\
(2011 HR)  & 1.25 & 0.47 & 50.8 & 20.09 & 0.22 & 42$\pm$6 & 1.35 & 202 (-40/+88) & 77---366 & 0.40 (-0.21/+0.24) & 0.04---0.70 \\
(2012 AS10)  & 1.56 & 1.20 & 40.5 & 17.59 & 0.22 & 44$\pm$6 & 1.25 & 750 (-180/+420) & 330---1900 & 0.29 (-0.17/+0.23) & 0.02---0.64 \\
(2014 MQ18)  & 1.57 & 0.92 & 38.1 & 16.15 & 0.22 & 4023$\pm$60 & 1.22 & 4800 (-1400/+3200) & 2400---14000 & 0.03 (-0.02/+0.03) & 0.00---0.10 \\
152560 (1991 BN) & 1.33 & 0.55 & 43.6 & 19.32 & 0.22 & 62$\pm$7 & 1.29 & 290 (-60/+140) & 120---570 & 0.38 (-0.21/+0.25) & 0.04---0.71 \\
337053 (1996 XW1) & 1.10 & 0.44 & 67.1 & 19.22 & 0.22 & 105$\pm$9 & 1.51 & 296 (-52/+106) & 100---393 & 0.41 (-0.20/+0.21) & 0.06---0.59 \\
312956 (1997 CZ3) & 1.04 & 0.17 & 74.8 & 19.80 & 0.22 & 3386$\pm$54 & 1.59 & 580 (-120/+230) & 220---780 & 0.06 (-0.03/+0.04) & 0.01---0.13 \\
53430 (1999 TY16) & 1.53 & 1.02 & 41.0 & 17.01 & 0.22 & 348$\pm$17 & 1.25 & 1580 (-430/+950) & 740---4220 & 0.11 (-0.07/+0.10) & 0.01---0.33 \\
331509 (1999 YA) & 1.41 & 0.66 & 41.1 & 18.45 & 0.22 & 104$\pm$9 & 1.25 & 490 (-110/+260) & 200---1170 & 0.31 (-0.18/+0.22) & 0.03---0.64 \\
68063 (2000 YJ66) & 2.18 & 1.51 & 24.1 & 15.86 & 0.22 & 61$\pm$7 & 1.10 & 2100 (-700/+1700) & 1100---9800 & 0.18 (-0.13/+0.22) & 0.01---0.63 \\
159533 (2001 HH31) & 1.75 & 1.08 & 31.9 & 17.97 & 0.22 & 37$\pm$6 & 1.17 & 710 (-200/+460) & 350---2390 & 0.22 (-0.14/+0.22) & 0.01---0.64 \\
286079 (2001 TW1) & 1.32 & 0.52 & 44.5 & 19.41 & 0.22 & 64$\pm$7 & 1.28 & 280 (-60/+130) & 110---580 & 0.39 (-0.21/+0.25) & 0.04---0.70 \\
(2002 VO85)  & 1.11 & 0.19 & 53.0 & 21.80 & 0.30 & 149$\pm$11 & 1.37 & 113 (-24/+50) & 45---189 & 0.26 (-0.14/+0.19) & 0.03---0.59 \\
(2002 VV17)  & 1.08 & 0.15 & 60.4 & 20.18 & 0.22 & 803$\pm$26 & 1.44 & 229 (-43/+91) & 80---332 & 0.28 (-0.14/+0.16) & 0.04---0.49 \\
280244 (2002 WP11) & 1.58 & 0.82 & 34.4 & 18.45 & 0.22 & 59$\pm$7 & 1.19 & 540 (-140/+320) & 240---1510 & 0.25 (-0.16/+0.22) & 0.01---0.64 \\
(2004 JR)  & 1.22 & 0.76 & 56.1 & 19.03 & 0.22 & 234$\pm$14 & 1.40 & 710 (-170/+340) & 300---1360 & 0.09 (-0.05/+0.07) & 0.01---0.22 \\
351508 (2005 RN33) & 1.37 & 0.55 & 40.4 & 19.89 & 0.22 & 47$\pm$6 & 1.25 & 250 (-60/+130) & 110---580 & 0.30 (-0.17/+0.22) & 0.02---0.63 \\
(2005 VC2)  & 1.23 & 0.40 & 49.3 & 18.07 & 0.22 & 29171$\pm$204 & 1.33 & 3600 (-900/+1900) & 1600---7400 & 0.01 (-0.00/+0.01) & 0.00---0.03 \\
(2005 XT77)  & 1.05 & 0.13 & 69.1 & 21.20 & 0.30 & 177$\pm$12 & 1.52 & 112 (-19/+38) & 37---137 & 0.46 (-0.21/+0.23) & 0.07---0.66 \\
417949 (2007 TB23) & 1.40 & 0.62 & 40.3 & 19.03 & 0.22 & 37$\pm$6 & 1.24 & 290 (-60/+130) & 110---630 & 0.51 (-0.28/+0.31) & 0.04---0.80 \\
(2007 WB5)  & 1.14 & 0.49 & 62.8 & 19.41 & 0.22 & 93$\pm$9 & 1.46 & 300 (-60/+120) & 110---460 & 0.33 (-0.16/+0.19) & 0.05---0.53 \\
(2009 FF19)  & 1.03 & 0.05 & 72.9 & 21.70 & 0.30 & 1938$\pm$41 & 1.56 & 136 (-26/+49) & 47---176 & 0.20 (-0.10/+0.13) & 0.03---0.38 \\
(2009 KN4)  & 1.09 & 0.52 & 67.7 & 18.55 & 0.22 & 365$\pm$18 & 1.51 & 590 (-120/+240) & 210---870 & 0.19 (-0.10/+0.12) & 0.03---0.35 \\
407338 (2010 RQ30) & 1.18 & 0.29 & 49.1 & 18.55 & 0.22 & 1228$\pm$32 & 1.33 & 530 (-120/+250) & 210---990 & 0.23 (-0.13/+0.16) & 0.02---0.46 \\
(2010 VF1)  & 1.14 & 0.28 & 57.0 & 20.70 & 0.30 & 241$\pm$14 & 1.40 & 229 (-49/+104) & 90---388 & 0.17 (-0.09/+0.13) & 0.01---0.42 \\
436761 (2012 DN) & 1.31 & 0.85 & 50.8 & 18.55 & 0.22 & 1015$\pm$29 & 1.35 & 1730 (-450/+950) & 790---3730 & 0.02 (-0.01/+0.02) & 0.00---0.07 \\
(2012 RG3)  & 1.07 & 0.24 & 69.9 & 18.26 & 0.22 & 1332$\pm$33 & 1.54 & 550 (-100/+200) & 180---720 & 0.29 (-0.14/+0.16) & 0.05---0.46 \\
(2012 UA34)  & 1.09 & 0.38 & 68.4 & 19.89 & 0.22 & 71$\pm$8 & 1.53 & 214 (-37/+74) & 71---268 & 0.43 (-0.19/+0.21) & 0.07---0.59 \\
(2013 QK48)  & 1.22 & 0.52 & 55.2 & 18.65 & 0.22 & 242$\pm$14 & 1.38 & 510 (-110/+230) & 200---910 & 0.24 (-0.12/+0.15) & 0.03---0.44 \\
(2014 NF64)  & 1.29 & 0.83 & 51.8 & 18.65 & 0.22 & 146$\pm$11 & 1.35 & 660 (-160/+340) & 280---1330 & 0.14 (-0.08/+0.11) & 0.01---0.34 \\
(2014 OL339)  & 1.05 & 0.12 & 67.7 & 23.00 & 0.30 & 71$\pm$8 & 1.51 & 58 (-11/+22) & 21---83 & 0.33 (-0.16/+0.19) & 0.04---0.57 \\
(2014 QW296)  & 1.55 & 0.81 & 36.0 & 19.13 & 0.22 & 63$\pm$7 & 1.20 & 520 (-150/+320) & 250---1470 & 0.15 (-0.09/+0.14) & 0.01---0.48 \\
136618 (1994 CN2) & 1.12 & 0.60 & 63.5 & 17.11 & 0.22 & 339$\pm$18 & 1.48 & 730 (-130/+270) & 240---970 & 0.47 (-0.22/+0.23) & 0.07---0.61 \\
137427 (1999 TF211) & 2.27 & 2.12 & 26.3 & 15.47 & 0.22 & 46$\pm$6 & 1.11 & 2900 (-1000/+2500) & 1600---13500 & 0.14 (-0.10/+0.18) & 0.01---0.54 \\
54509 YORP (2000 PH5) & 1.04 & 0.05 & 58.5 & 22.90 & 0.30 & 282$\pm$16 & 1.42 & 47 (-8/+15) & 16---60 & 0.56 (-0.25/+0.27) & 0.09---0.71 \\
228368 (2000 WK10) & 1.12 & 0.52 & 64.2 & 18.65 & 0.22 & 206$\pm$13 & 1.48 & 470 (-90/+180) & 170---680 & 0.28 (-0.14/+0.16) & 0.04---0.47 \\
(2002 AQ2)  & 1.09 & 0.42 & 68.5 & 18.93 & 0.22 & 364$\pm$18 & 1.52 & 490 (-100/+190) & 170---690 & 0.20 (-0.10/+0.12) & 0.03---0.37 \\
(2012 GG1) & 1.18 & 0.71 & 58.6 & 18.93 & 0.22 & 219$\pm$14 & 1.43 & 620 (-140/+270) & 250---1080 & 0.12 (-0.06/+0.09) & 0.01---0.28 \\
\enddata

\tablecomments{
The columns are as follows: target name;
heliocentric and Spitzer-centric distances; phase
angle of observations (in Spitzer-centric
reference frame);
corrected $H$~magnitude and applied uncertainty
(see Section~\ref{Hmag});
in-band measured fluxes at
4.5~microns, where the uncertainties do not
include the 5\% calibration uncertainty;
nominal $\eta$ value applied in the modeling,
as calculated from phase angle (see description
in text);
and derived albedos and diameters,
where 1$\sigma$ uncertainties are given 
in parentheses and 3$\sigma$ ranges are
listed.
Note that 
the in-band fluxes listed
here must be color-corrected as
described in the text in order to 
carry out the thermal modeling and
derive diameters and albedos.
In some cases the albedo errors formally include
zero, which should be interpreted as indicating
that the minimum albedo is a small non-zero number.
Further information about the observations and model 
results (including
AORKey, which is a unique Spitzer identifier for
each observation; date and time of the observations;
details of the exposure times; and
uncertainties in $\eta$)
is available
from our online database; see Table~\ref{webpage}
for more details.
}
\end{deluxetable}

\begin{deluxetable}{llll}
\tabletypesize{\small}
\rotate
\tablecaption{Properties listed in our online table \label{webpage}}
\tablewidth{0pt}
\tablehead{}
\startdata
Object number & Object name & Object designation \\
Semi-major axis & Eccentricity & Inclination \\
Long.\ asc.\ node $\Omega$ & Arg.\ perihelion $\omega$ & Orbital period \\
Observed RA & Observed Dec & 3$\sigma$ RA Uncert.\ & 3$\sigma$ Dec Uncert. \\
V magnitude & Heliocentric distance & Spitzer-centric distance \\
Phase angle & Solar elong.\ angle & Galactic latitude & Galactic longitude \\
Date of observation & JD of observation & AORKey \\
Frame time & Total integration time & Elapsed time & Notes about the observation or data quality \\
CH2 flux & CH2 error & CH2 SNR \\
$H$ magnitude & Uncertainty on $H$ magnitude & G (slope parameter) \\
$\eta$ & $\eta$: $\pm$1$\sigma$ & $\eta$: $\pm$3$\sigma$ \\
Diameter & Diameter: $\pm$1$\sigma$ & Diameter: $\pm$3$\sigma$ \\
Albedo & Albedo: $\pm$1$\sigma$ & Albedo: $\pm$3$\sigma$ \\
\enddata
\tablecomments{These quantities and their units are
fully defined
at {\tt nearearthobjects.nau.edu} .}
\end{deluxetable}







\end{document}